\documentclass[aps, prl, twocolumn, superscriptaddress, reprint]{revtex4-1}

\usepackage{siunitx}
\usepackage{color}
\usepackage{upgreek}
\usepackage{graphicx}
\usepackage{dcolumn}
\usepackage{bm}
\begin{document}

\title{Time refraction of spin waves} 

\author{K. Schultheiss}\email{k.schultheiss@hzdr.de}
\affiliation{Helmholtz-Zentrum Dresden-Rossendorf, Institut f\"{u}r Ionenstrahlphysik und Materialforschung, 01328 Dresden, Germany}

\author{N. Sato}
\affiliation{Helmholtz-Zentrum Dresden-Rossendorf, Institut f\"{u}r Ionenstrahlphysik und Materialforschung, 01328 Dresden, Germany}

\author{P. Matthies}
\affiliation{Helmholtz-Zentrum Dresden-Rossendorf, Institut f\"{u}r Ionenstrahlphysik und Materialforschung, 01328 Dresden, Germany}
\affiliation{Fakult\"{a}t Physik, Technische Universit\"{a}t Dresden, 01062 Dresden, Germany}

\author{L. K\"{o}rber}
\affiliation{Helmholtz-Zentrum Dresden-Rossendorf, Institut f\"{u}r Ionenstrahlphysik und Materialforschung, 01328 Dresden, Germany}
\affiliation{Fakult\"{a}t Physik, Technische Universit\"{a}t Dresden, 01062 Dresden, Germany}

\author{K. Wagner}
\affiliation{Helmholtz-Zentrum Dresden-Rossendorf, Institut f\"{u}r Ionenstrahlphysik und Materialforschung, 01328 Dresden, Germany}

\author{T. Hula}
\affiliation{Helmholtz-Zentrum Dresden-Rossendorf, Institut f\"{u}r Ionenstrahlphysik und Materialforschung, 01328 Dresden, Germany}
\affiliation{Institut f\"{u}r Physik, Technische Universit\"{a}t Chemnitz, 09107 Chemnitz, Germany}

\author{O. Gladii}
\affiliation{Helmholtz-Zentrum Dresden-Rossendorf, Institut f\"{u}r Ionenstrahlphysik und Materialforschung, 01328 Dresden, Germany}

\author{J. E. Pearson}
\affiliation{Materials Science Division, Argonne National Laboratory, Argonne, Illinois 60439, USA}

\author{A. Hoffmann}
\altaffiliation[Permanent address:]{ Department of Materials Science and Engineering, University of Illinois at Urbana-Champaign, Urbana, IL 61801, USA}
\affiliation{Materials Science Division, Argonne National Laboratory, Argonne, Illinois 60439, USA}

\author{M. Helm}
\affiliation{Helmholtz-Zentrum Dresden-Rossendorf, Institut f\"{u}r Ionenstrahlphysik und Materialforschung, 01328 Dresden, Germany}
\affiliation{Fakult\"{a}t Physik, Technische Universit\"{a}t Dresden, 01062 Dresden, Germany}

\author{J. Fassbender}
\affiliation{Helmholtz-Zentrum Dresden-Rossendorf, Institut f\"{u}r Ionenstrahlphysik und Materialforschung, 01328 Dresden, Germany}
\affiliation{Fakult\"{a}t Physik, Technische Universit\"{a}t Dresden, 01062 Dresden, Germany}

\author{H. Schultheiss}
\affiliation{Helmholtz-Zentrum Dresden-Rossendorf, Institut f\"{u}r Ionenstrahlphysik und Materialforschung, 01328 Dresden, Germany}
\affiliation{Fakult\"{a}t Physik, Technische Universit\"{a}t Dresden, 01062 Dresden, Germany}

\date{\today}

\begin{abstract}

We present an experimental study of time refraction of spin waves propagating in microscopic waveguides under the influence of time-varying magnetic fields. Using space- and time-resolved Brillouin light scattering microscopy, we demonstrate that the broken translational symmetry along the time coordinate can be used to in- or decrease the energy of spin waves during their propagation. This allows for 
a broadband and controllable shift of the spin-wave frequency. Using an integrated design of spin-wave waveguide and microscopic current line for the generation of strong, nanosecond-long, magnetic field pulses, a conversion efficiency up to 39\% of the carrier spin-wave frequency is achieved,  Êsignificantly larger compared to photonic systems. Given the strength of the magnetic field pulses and its strong impact on the spin-wave dispersion relation, the effect of time refraction can be quantified on a length scale comparable to the spin-wave wavelength. Furthermore, we utilize time refraction to excite spin-wave bursts with pulse durations in the nanosecond range and a frequency shift depending on the pulse polarity.

\end{abstract}


\maketitle 

The manipulation of waves can be achieved by modulating the properties of the medium in which the wave propagates either in space or in time. Changes along the space coordinates affect the momentum of the wave due to the lost translational invariance of space but conserve the energy of the wave. Well known examples are the diffraction of waves at opaque obstacles or the refraction of light according to Snell's law. While these phenomena are classified by the term \textit{space refraction}, its counterpart is known as \textit{time refraction} which conserves the momentum of the wave but changes its energy due to the lost  translational invariance of time \cite{Mendonca2000}. Key requirement for time refraction is to temporarily change the medium parameters, which affect the dispersion relation of the wave, in an adiabatic manner, i.e., slower than the eigenfrequency of the wave, but faster than its lifetime. Recently, Zhou and coworkers reported on the experimental observation of time refraction of photons with an unprecedented high frequency conversion efficiency of about 6\%  at a photon wavelength of \SI{1240}{\nano\meter} in a slab of indium tin oxide with subwavelength thickness \cite{Boyd2020}. In photonic systems, it is challenging to find  materials in which the relative change of the refractive index in time is large enough to detect a sizable change in frequency and still keep the interaction length of photons and refractive medium on the order of the photon wavelength or even smaller. In ferromagnets, however, which host spin waves as collective excitations of the spin system, the strong coupling of spin waves to the magnetic field and their rather small phase velocity compared to the speed of light allows for even higher relative frequency shifts and a full spatio-temporal study of time refraction.

The dispersion relation of spin waves depends strongly on the effective magnetic field, the magnitude and direction of the magnetization and the thickness and lateral dimensions of the spin-wave waveguide \cite{Kalinikos1986}. Typically, spin waves have frequencies in the gigahertz range and their phase- and group-velocities are comparable to the speed of sound in solids. Space refraction of spin waves was already demonstrated in the last two decades for spatially inhomogeneous magnetic fields \cite{Chumak2009}, changes of the waveguide width \cite{Demidov2011}, spatial variations of the magnetization generated by heat gradients \cite{Obry}, directional changes of the magnetization \cite{Vogt2014, Kim2008} and thickness variations along the spin-wave propagation path \cite{Back2016, Mendach2016}. In all these studies, changes of the spin-wave momenta were reported under conservation of the spin-wave energy. However, there are only  few reports on time refraction of spin waves in macroscopic yttrium-iron-garnet waveguides dating back up to 50 years \cite{Rezende1967, Zapp1967, Fetisov}. Given the lack of spatially resolved methods for mapping spin-wave propagation at that time, these studies could only analyze the change of the spin-wave energy after propagation over several millimeters which limited the speed for changing the magnetic field and the resulting frequency shifts to the megahertz regime. 

\begin{figure}[t!]
  \begin{center}
    \includegraphics[width=8.5cm]{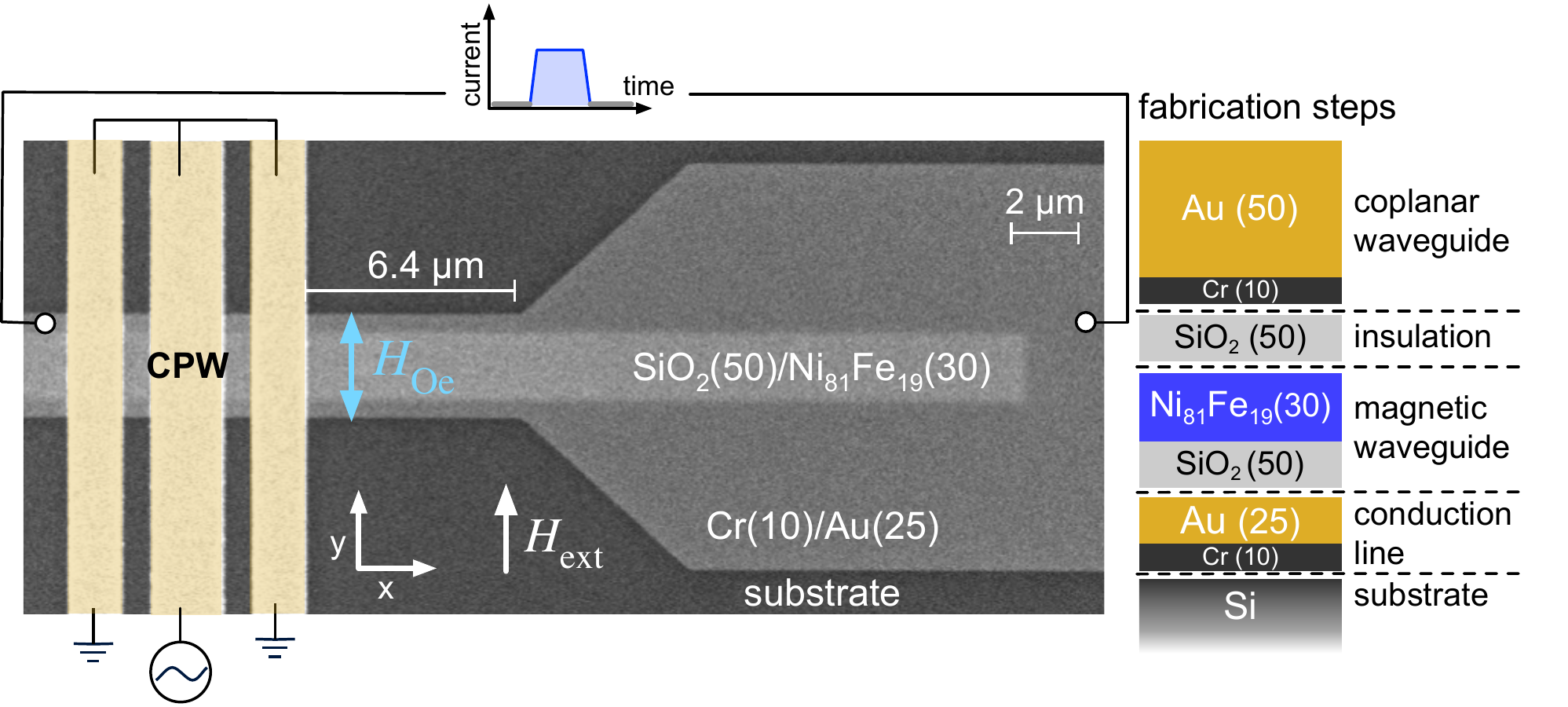}
      \end{center}
  \caption{Scanning-electron-microscopy (SEM) image of the investigated structure. A Ni$_{81}$Fe$_{19}$ waveguide (width $w=\SI{2}{\micro\meter}$, thickness $t= \SI{30}{\nano\meter}$) is fabricated on top of a Cr(\SI{10}{\nano\meter})/Au(\SI{25}{\nano\meter}) conduction line ($w=\SI{3.2}{\micro\meter}$), insulated by \SI{50}{\nano\meter} SiO$_{2}$. Spin waves are excited by microwave currents flowing through a coplanar waveguide (CPW), which is shorted outside the area of the SEM image. Current pulses injected into the Au conduction line generate pulsed Oersted fields ${\bm {H}_{\rm{Oe}}}$. An external magnetic field $\bm {H}_{\rm{ext}}$ is applied along the $+y$ direction. The fabrication steps highlight the insulation layers, which are not visible in the SEM image. }
    \label{fig1}
\end{figure}

In this Letter, we report on micron sized spin-wave waveguides with integrated current lines for generating large and rapidly changing magnetic fields using current pulses. For the first time, we present a full spatio-temporal quantification of time refraction of spin waves using time-resolved Brillouin light scattering microscopy (TR-$\upmu$BLS) \cite{BLS}. We demonstrate a relative change of the spin-wave frequency up to 39\% over propagation distances comparable to the spin-wave wavelength. Depending on the polarity of the $dc$ pulses the spin-wave energy can be increased or decreased within a few nanoseconds. 

The experimental layout is shown in Fig.~\ref{fig1}. We pattern a ferromagnetic Ni$_{81}$Fe$_{19}$ waveguide (width $w=\SI{2}{\micro\meter}$, thickness $t= \SI{30}{\nano\meter}$) on top of a  Cr(\SI{10}{\nano\meter})/Au(\SI{25}{\nano\meter}) conduction line ($w=\SI{3.2}{\micro\meter}$), as can be seen in the scanning electron micrograph in Fig.~\ref{fig1}. Both layers are insulated by \SI{50}{\nano\meter} SiO$_{2}$. A current pulse flowing in the conduction line generates a pulsed Oersted field ${\bm {H}_{\rm{Oe}}}$, which is aligned either parallel or antiparallel to the direction of an external magnetic field ${\bm {H}_{\rm{ext}}}$, depending on the polarity of the current pulse. This allows for a rapid change of the total magnetic field inside the Ni$_{81}$Fe$_{19}$ waveguide. Note that the broadening of the Au conduction line in $y$-direction is not relevant for this study since all measurements were performed in the \SI{6.4}{\micro\meter} long part before the broadening begins.

\begin{figure}
  \begin{center}
      \includegraphics[width=8.5cm]{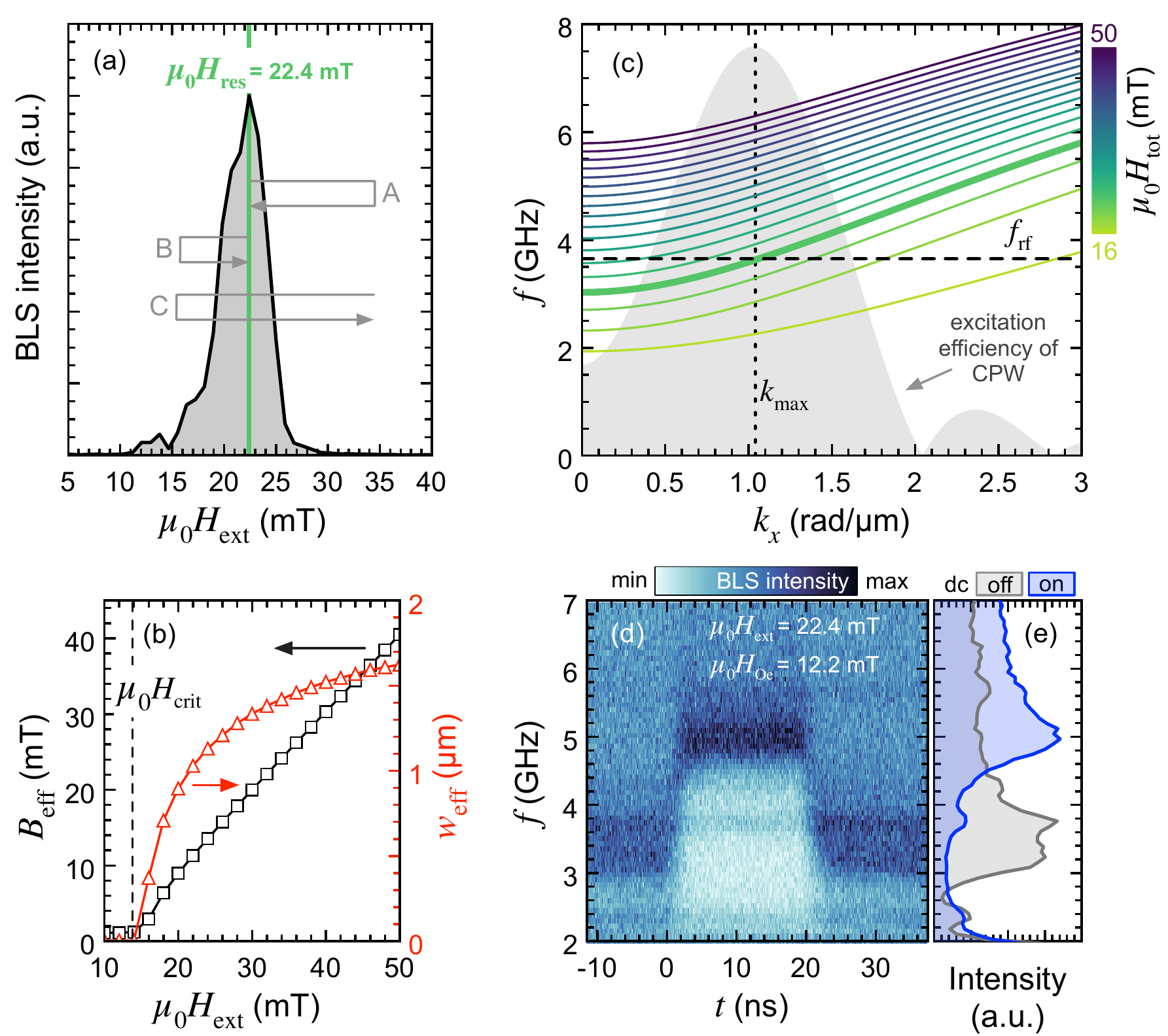}
  \end{center}
  \caption{(a) BLS intensity measured for spin waves excited at $f_{\rm{rf}}=\SI{3.65}{\giga\hertz}$ as a function of the externally applied field, yielding the resonance field $\mu_{0}H_{\rm{res}}=\SI{22.4}{\milli\tesla}$. The arrows A to C depict different sequences of the time varying magnetic fields. (b) Effective field $B_{\rm{eff}}$ and effective localization width $w_{\rm{eff}}$ determined from micromagnetic simulations for different external magnetic fields. (c) Dispersion relations calculated for different external magnetic fields, considering the resulting $B_{\rm{eff}}$ and $w_{\rm{eff}}$ as in b. The dotted vertical line indicates the wave vector $k_{\rm{max}}$ which is excited most efficiently by the CPW. The horizontal dashed line indicates the excitation frequency $f_{\rm{rf}}$, fixed in all measurements at \SI{3.65}{\giga\hertz}. (d) BLS measurement of the thermal spin-wave intensity as a function of frequency and time with $\mu_{0}H_{\rm{ext}}=\SI{22.4}{\milli\tesla}$ and a \SI{20}{\nano\second} long $dc$ pulse, resulting in  $\mu_{0}H_{\rm{Oe}}=\SI{12.2}{\milli\tesla}$. The BLS intensity is color coded. (e) Frequency spectra integrated in time windows when the $dc$ pulse is on and off, respectively.}
    \label{fig2}
\end{figure}

A microwave current flowing through a coplanar waveguide (CPW, see Fig.~\ref{fig1}) excites spin waves with a fixed frequency  $f_{\rm{rf}}=\SI{3.65}{\giga\hertz}$ and a preferential wave vector $k_{\rm{max}}=\SI{1.04}{rad/\micro\meter}$. To determine the resonance field for these specific excitation conditions, we use Brillouin light scattering microscopy ($\upmu$BLS) and measure the spin-wave intensity in \SI{1}{\micro\meter} distance to the CPW as a function of the external magnetic field $H_{\rm{ext}}$, without applying any $dc$ current pulses. As can be seen from the results in Fig.~\ref{fig2}(a), spin waves are excited most efficiently at the resonance field $\mu_{0}H_{\rm{res}}=\SI{22.4}{\milli\tesla}$. 

In order to compare this to the dispersion relation, we need to take into account the strong influence of the demagnetizing field and the quantization of the spin-wave wave vector across the waveguide width. Therefore, we simulate the static magnetization inside the waveguide for external fields between \SI{10}{\milli\tesla} and \SI{50}{\milli\tesla} using MuMax$^3$\cite{MuMax, dispersion}. Black squares in Fig.~\ref{fig2}(b) show the effective field $B_\mathrm{eff}$, which is given by the external field reduced by the demagnetizing field and is determined in the center of the waveguide. The effective quantization width $w_\mathrm{eff}$ is given by red triangles in Fig.~\ref{fig2}(b) and is determined from the positions at which the $y$-component of the magnetization reaches 99\% of the saturation magnetization $m_{y}= 0.99  M_{\rm{S}}$. For $H_{\rm{ext}}\leq H_{\rm{crit}}=\SI{14}{\milli\tesla}$, the external field is too weak to compensate the demagnetizing fields so that the magnetization is aligned predominantly parallel to the long axis of the waveguide. 

Based on the results for $B_{\rm{eff}}$ and $w_{\rm{eff}}$, we plot dispersion relations for various magnetic fields $\SI{16}{\milli\tesla} \leq \mu_{0}H_{\rm{tot}}\leq\SI{50}{\milli\tesla}$ in Fig.~\ref{fig2}(c). The calculations follow the formalism of Kalinikos and Slavin\cite{Kalinikos1986}, taking into account the quantization of the wave vector across the waveguide width $k_y=\pi/w_\mathrm{eff}$ and using the same material parameters as in the micromagnetic simulations \cite{dispersion}. The excitation conditions for spin waves in our system ($f_{\rm{rf}}=\SI{3.65}{\giga\hertz}$, $k_{\rm{max}}=\SI{1.04}{rad/\micro\meter}$) match the dispersion calculated for $\mu_{0}H_{\rm{ext}}=\SI{22}{\milli\tesla}$ [thicker green line in Fig.~\ref{fig2}(c)], which is in good agreement with the measured resonance field $\mu_{0}H_{\rm{res}}=\SI{22.4}{\milli\tesla}$. 

As a first step, we demonstrate the influence of the pulsed Oersted fields in TR-$\mu$BLS measurements of the thermal spin-wave signal, when no microwaves are applied to the CPW. Figure~\ref{fig2}(d) shows the temporal evolution of thermal spin-wave spectrum for $\mu_0 H_{\rm{ext}}=\SI{22.4}{\milli\tesla}$ with the BLS intensity color coded. Up to $t=\SI{0}{\nano\second}$, only the external field determines the spin-wave frequencies with a minimum at \SI{3.0}{\giga\hertz}. At $t=\SI{0}{\nano\second}$, a current pulse is applied to the Au conduction line with \SI{20}{\nano\second} duration, \SI{3}{\nano\second} rise and fall time, and $I_{\rm{dc}}=\SI{64.2}{\milli\ampere}$ amplitude. Inside the Ni$_{81}$Fe$_{19}$ waveguide, this generates an Oersted field $\mu_0 H_{\rm{Oe}}=\SI{12.2}{\milli\tesla}$, which is aligned parallel to the external field and, thus, increases the overall magnetic field. Hence, in this \SI{20}{\nano\second} window, the spin-wave spectrum is shifted to higher frequencies with a rising flank around  \SI{4.6}{\giga\hertz}. To better visualize this shift, we integrate the BLS intensity inside (dc on) and outside (dc off) the \SI{20}{\nano\second} window and plot both spectra in Fig.~\ref{fig2}(e). The frequencies of the rising flanks without (\SI{3.0}{\giga\hertz}) and with (\SI{4.6}{\giga\hertz}) additional $dc$ pulse are in excellent agreement with the onsets of the dispersion relation at  $\mu_0 H_{\rm{tot}}=\SI{22.4}{\milli\tesla}$ and $\mu_0 H_{\rm{tot}}=\SI{34.6}{\milli\tesla}$, respectively. 

Now, we study how spin waves adapt to temporal changes of the magnetic field if a continuous microwave signal is applied to the CPW at $f_{\rm{rf}}=\SI{3.65}{\giga\hertz}$. Therefore, we distinguish different sequences of  time varying magnetic fields: the externally applied magnetic field matches the resonance condition so that any additional $dc$ pulse shifts  spin waves out of resonance [see arrows A and B in Fig.~\ref{fig2}(a)] or the initial external magnetic field is larger than the resonance field, and the $dc$ pulse is injected so that the associated Oersted field  reduces the overall magnetic field to cross the resonance condition [see arrow C in Fig.~\ref{fig2}(a)].  

\begin{figure}
  \begin{center}
    \includegraphics[width=8.5Cm]{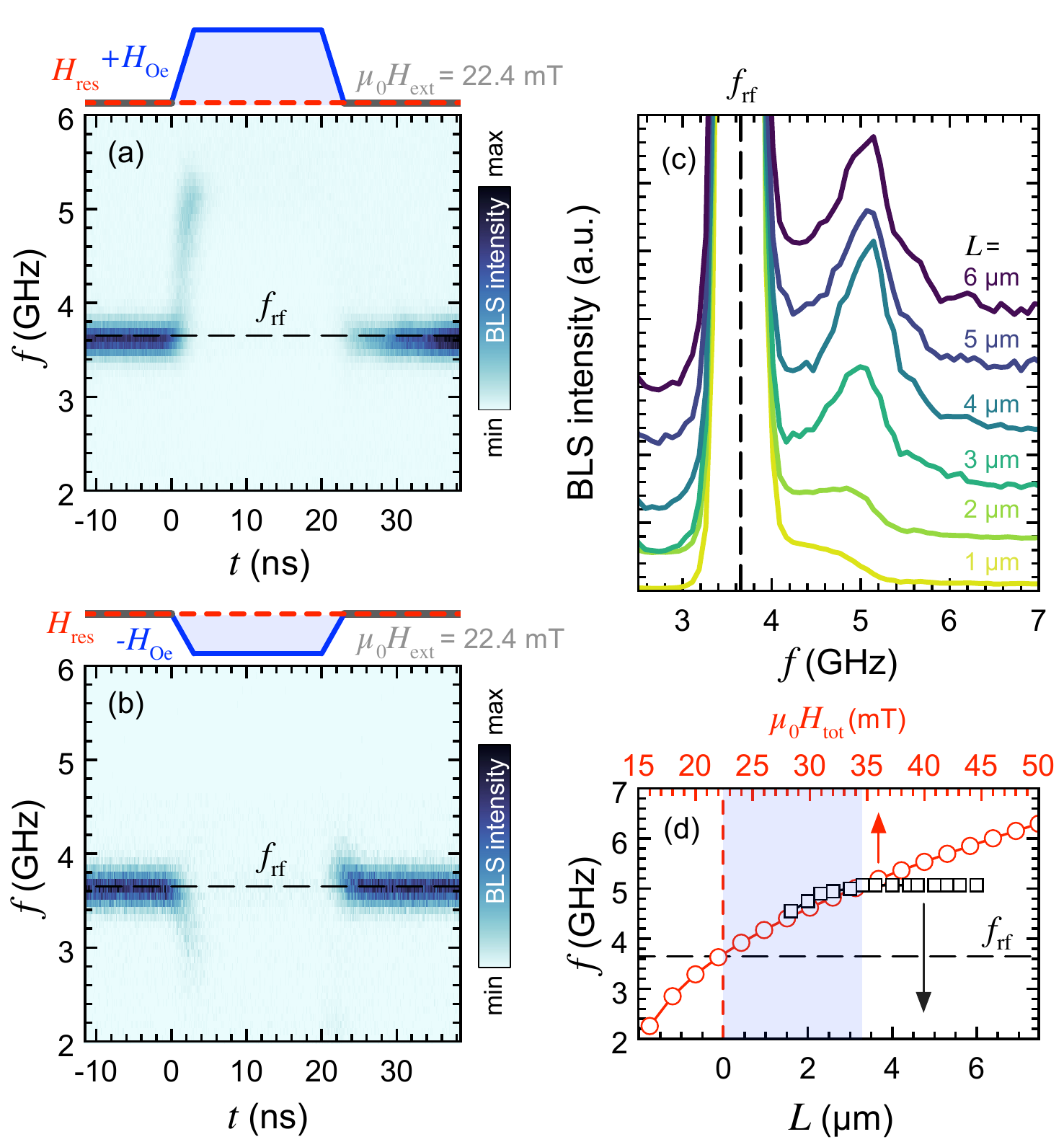}
  \end{center}
  \caption{TR-BLS signal measured at $L=\SI{4}{\micro\meter}$ for (a) ${\bm {H}_{\rm{Oe}}} \uparrow\uparrow {\bm {H}_{\rm{ext}}}$ with $\mu_{0}H_{\rm{Oe}}=\SI{12.2}{\milli\tesla}$, and (b) ${\bm {H}_{\rm{Oe}}} \uparrow\downarrow {\bm {H}_{\rm{ext}}}$ with $\mu_{0}H_{\rm{Oe}}=\SI{6.8}{\milli\tesla}$. The excitation was fixed at $f_{\rm{rf}}=\SI{3.65}{\giga\hertz}$ (horizontal dashed lines). (c)  Spin-wave intensities integrated over all times and measured at different distances $L$ to the edge of the CPW. Each graph is normalized to the intensity of its resonance peak and shifted vertically for clarity. (d) Frequencies measured for the shifted spin-wave peak as function of $L$ (black squares) compared to frequencies calculated from the dispersion relation for the wave vector $k_{\rm{max}}=\SI{1.04}{rad/\micro\meter}$ as function of the total magnetic field (red circles). The light blue area highlights the field range covered by the $dc$ pulse. }
    \label{fig3}
\end{figure}

We start with the case when the external field is set to the resonance field $\mu_{0}H_{\rm{ext}}=\mu_{0}H_{\rm{res}}=\SI{22.4}{\milli\tesla}$ and the  $dc$ pulse generates an additional Oersted field $\mu_{0}H_{\rm{Oe}}=\SI{12.2}{\milli\tesla}$ parallel to the external field [arrow A in Fig.~\ref{fig2}(a)]. As in the thermal measurement, the pulse duration is \SI{20}{\nano\second} with \SI{3}{\nano\second} rise and fall time. Figure~\ref{fig3}(a) shows the corresponding intensity plot of the $\mu$BLS signal which was measured in $L=\SI{4}{\micro\meter}$ distance from the edge of the CPW. Up to $t= \SI{0}{\nano\second}$, i.e., before injecting the $dc$ pulse, a strong BLS signal is detected at \SI{3.65}{\giga\hertz}, as expected for resonant excitation. 

During the \SI{3}{\nano\second} rise time of the $dc$ pulse, we detect a rapid increase of the spin-wave frequency. As explained before, this frequency shift is due to the parallel alignment of external field and Oersted field ${\bm {H}_{\rm{Oe}}} \uparrow\uparrow {\bm {H}_{\rm{ext}}}$ which causes an increase of the total magnetic field $H_{\rm{tot}}$ and shifts the dispersion to higher frequencies. However, unlike in the case of thermal spin waves, the spin-wave signal vanishes after the $dc$ pulse reaches its maximum. Only as long as $\mu_{0}H_{\rm{tot}}=\mu_{0}H_{\rm{res}}=\SI{22.4}{\milli\tesla}$, spin waves are resonantly excited by the coplanar waveguide. With the additional Oersted field, $H_{\rm{tot}}>H_{\rm{res}}$ and no more spin waves can be excited at $f_\mathrm{rf}=\SI{3.65}{\giga\hertz}$. Only those spin waves, that were excited before the $dc$ pulse sets in, adjust their frequency due to time refraction while they propagate to the point of detection. Naturally, after the $dc$ pulse is switched off, the total magnetic field  matches again the resonance condition. New spin waves can be excited by the coplanar waveguide and propagate along the waveguide.  

If we keep the external field to match the resonance condition $\mu_{0}H_{\rm{ext}}=\mu_{0}H_{\rm{res}}=\SI{22.4}{\milli\tesla}$ but reverse the direction of the Oersted field [arrow B in Fig.~\ref{fig2}(a)] with reduced amplitude $\mu_{0}H_{\rm{Oe}}=\SI{6.8}{\milli\tesla}$, the result is similar. Now, ${\bm {H}_{\rm{Oe}}} \uparrow\downarrow {\bm {H}_{\rm{ext}}}$ so that the total field is reduced during the $dc$ pulse and the spin-wave frequency shifts to lower values during the rise time of the pulse, as can be seen in Fig.~\ref{fig3}(b). Note that this results in $H_{\rm{tot}}< H_{\rm{crit}}$ [see Fig.~\ref{fig2}(b)] which does not suffice to saturate even parts of the waveguide along its short axis. Hence, spin-wave propagation is less efficient and the shifted signal in Fig.~\ref{fig3}(b) is much weaker compared to Fig.~\ref{fig3}(a).

So far, we  detected  spin waves only in one fixed distance to the edge of the CPW. Now, we analyze the amplitude of the frequency shift as a function of the propagation distance $L$. In Fig.~\ref{fig3}(c), we plot the $\upmu$BLS intensity integrated over time that was measured at different $L$ for ${\bm{H}_{\rm{Oe}}} \uparrow\uparrow {\bm{H}_{\rm{ext}}}$ with $\mu_{0}H_{\rm{ext}}=\SI{22.4}{\milli\tesla}$ and $\mu_{0}H_{\rm{Oe}}=\SI{12.2}{\milli\tesla}$. Each spectrum is normalized to the intensity of the resonance peak at \SI{3.65}{\giga\hertz}, which corresponds to spin waves excited at resonance. The second peak at higher frequencies resembles spin waves that adapted their frequencies to the time varying magnetic field. The frequencies of these shifted peaks are summarized by black squares in Fig.~\ref{fig3}(d) for all measured distances $L$. This frequency gradually increases until it reaches a plateau at $\Delta f=\SI{5.07}{\giga\hertz}$ for $L=\SI{3.3}{\micro\meter}$, which corresponds to 39\% relative frequency shift due to time refraction. This indicates that the Oersted field induced by the $dc$ pulse reaches its maximum of \SI{12.2}{\milli\tesla} when spin waves reach position $L=\SI{3.3}{\micro\meter}$.

To compare our data to the dispersion relation, red circles in Fig.~\ref{fig3}(d) show the frequencies that we calculated for $k_{\rm{max}}=\SI{1.04}{rad/\micro\meter}$ from  the dispersion relation in Fig.~\ref{fig2}(c). In order to match the different $x$-axes for both sets of data, we have to consider that when measuring directly at the antenna, i.e., at $L=\SI{0}{\micro\meter}$, the field did not change yet and coincides with the externally applied field $\mu_{0}H_{\rm{ext}}=\SI{22.4}{\milli\tesla}$. At $L=\SI{3.3}{\micro\meter}$, no shift is detected any more so that the maximum total field is reached $\mu_{0}H_{\rm{tot}}=\mu_{0}(H_{\rm{ext}}+H_{\rm{Oe}}) =\SI{34.6}{\milli\tesla}$. As can be seen, measured and calculated frequencies coincide, up to the point when the measured data reach the plateau.
 
\begin{figure}
  \begin{center}
    \includegraphics[width=85mm]{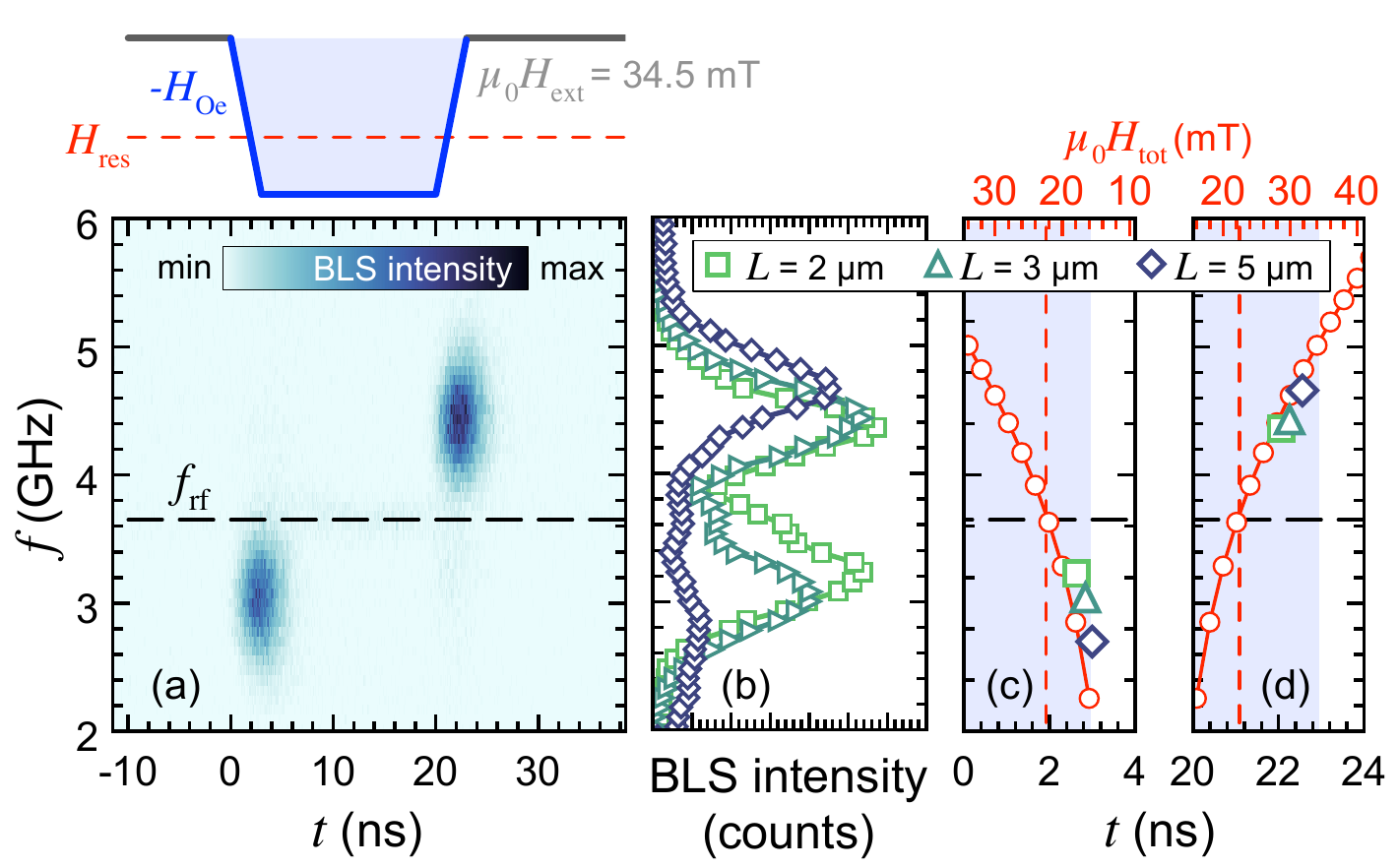}
  \end{center}
  \caption{(a) Time-resolved BLS signal measured at $L= \SI{3}{\micro\meter}$ for ${\bm{H}_{\rm{Oe}}} \uparrow\downarrow {\bm{H}_{\rm{ext}}}$  with $\mu_0 H_{\rm{ext}}=\SI{34.5}{\milli\tesla}$ and $\mu_0H_{\rm{Oe}}=\SI{19.0}{\milli\tesla}$. The excitation frequency was fixed to $f_{\rm{rf}}=\SI{3.65}{\giga\hertz}$ (horizontal dashed line).  (b) BLS spectra integrated over time for different $L$ with the same field configuration as in a. (c),(d) Frequencies and detection times of the spin-wave bursts measured at different $L$ for (c) the rising edge of the $dc$ pulse and (d) the falling edge. The \SI{3}{\nano\second} rise and fall time are highlighted in light blue.}
    \label{fig4}
\end{figure}

Now we consider the case when the external magnetic field is larger than the resonance field and the $dc$ pulse is injected to generate an Oersted field that reduces the effective field to cross the resonance condition [arrow C in Fig.~\ref{fig2}(a)].  Figure~\ref{fig4}(a) shows the BLS intensity for ${\bm{H}_{\rm{Oe}}} \uparrow\downarrow {\bm{H}_{\rm{ext}}}$  with $\mu_0 H_{\rm{ext}}=\SI{34.5}{\milli\tesla}$ and $\mu_0H_{\rm{Oe}}=\SI{19.0}{\milli\tesla}$. Only for two short moments in time, during the rise and fall time of the $dc$ pulse, spin-wave bursts are detected. Only at these times, the total field matches the resonance field  $\mu_0H_{\rm{tot}}=\mu_0H_{\rm{res}}=\SI{22.4}{\milli\tesla}$ and allows for the excitation of spin-wave bursts with $f_{\rm{rf}}=\SI{3.65}{\giga\hertz}$. However, the spin-wave bursts are not detected at \SI{3.65}{\giga\hertz} but at lower (higher) values at the rising (falling) edge of the $dc$ pulse, respectively. This is related to the still changing total magnetic field while the spin-wave bursts propagate. At the rising edge of the $dc$ pulse, the total field continues to reduce, leading to a downshift of the burst frequency. At the falling edge, the total field still increases, resulting in a higher average frequency of the spin-wave burst. 

To analyze the frequency shifts of the spin-wave bursts in more detail, Fig.~\ref{fig4}(b) shows the BLS intensity integrated over time and measured at different distances $L$ to the antenna. Moving away from the antenna, the frequency shift of the spin-wave burst increases, similar to our observations for resonant excitation discussed above. The larger the propagation distance, the longer the bursts propagate in the time-varying field and, hence, the more their frequencies are shifted. 

To compare this with calculations of the dispersion, the frequencies of the spin-wave bursts are plotted as a function of time in Fig.~\ref{fig4}(c) for the rising edge of the $dc$ pulse and in Fig.~\ref{fig4}(d) for the falling edge. Red circles show the frequencies that were calculated from the dispersion for $k_{\rm{max}}=\SI{1.04}{rad/\micro\meter}$ as a function of the total magnetic field (upper $x$-axis). To match both $x$-axes, we have to consider that for the rising edge of the $dc$ pulse, the total field at $t=\SI{0}{\nano\second}$ still matches the externally applied field: $\mu_0 H_{\rm{tot, \SI{0}{\nano\second}}}=\SI{34.5}{\milli\tesla}$. After \SI{3}{\nano\second} rise time, the pulse amplitude reaches its maximum which yields $\mu_0 H_{\rm{Oe}}=\SI{19.0}{\milli\tesla}$ and $\mu_0 H_{\rm{tot, \SI{3}{\nano\second}}}=\SI{15.5}{\milli\tesla}$. As can be seen in Fig.~\ref{fig4}(c), the measured frequencies of the spin-wave bursts nicely coincide with what is expected from theory. For the falling edge of the $dc$ pulse [Fig.~\ref{fig4}(d)], the axis of the total magnetic field is simply reversed, i.e., $\mu_0 H_{\rm{tot, \SI{20}{\nano\second}}}=\SI{15.5}{\milli\tesla}$ and $\mu_0 H_{\rm{tot, \SI{23}{\nano\second}}}=\SI{34.5}{\milli\tesla}$. Also here, experimental data  and theoretical calculations match well. 

In conclusion, we have experimentally quantified time refraction of spin waves in combined space- and time-resolved studies of spin-wave propagation in rapidly changing magnetic fields. We demonstrated the acceleration and deceleration of spin waves with frequencies in the GHz range on length scales comparable to the spin-wave wavelength. The prospect of manipulating spin-wave frequencies on such short time- and length-scales is of fundamental importance for hybrid quantum systems in which magnons, the quanta of spin waves, are considered as a quantum memory \cite{Nakamura2015} or a mediator for long distance entanglement of solid-state qubits \cite{Loss2013, Awschalom2017}.

\begin{acknowledgments}
This work was supported by the Deutsche Forschungsgemeinschaft within program SCHU 2922/1-1. K.S. acknowledges funding within the Helmholtz Postdoc Programme. Discussions with A.N. Slavin are gratefully acknowledged. Samples were prepared at the Argonne National Laboratory with thin film growth supported by the U.S. Department of Energy, Office of Science, Materials Science and Engineering Division. Lithography was performed at the Center for Nanoscale Materials, an Office of Science user facility, which was supported by the U.S. Department of Energy, Office of Science, Basic Energy Sciences, under Contract No. DE-AC02-06CH11357. 
\end{acknowledgments}



\begin{thebibliography}{99}
\bibitem{dispersion} The dispersion relation and micromagnetic simulations were performed using the following material parameters: saturation magnetization $M_{\rm{s}}=\SI{830}{\kilo\ampere/\meter}$, reduced gyromagnetic ratio $\gamma/2\pi =\SI{28}{\giga\hertz/\tesla}$, exchange constant $A=\SI{13}{\pico\joule/\meter}$.
\end{thebibliography}
\end{document}